\documentstyle[amssymb,prb,aps]{revtex}
\begin{document}
\title{Observation of the cluster spin-glass phase in La$_{2-x}$Sr$_{x}$CuO$_{4}$
by anelastic spectroscopy}
\author{F. Cordero}
\address{CNR, Area di Ricerca di Tor Vergata, Istituto di Acustica ``O.M. Corbino``,\\
Via del Fosso del Cavaliere 100, I-00133 Roma, and INFM, Italy}
\author{R. Cantelli, A. Paolone}
\address{Universit\`{a} di Roma ``La Sapienza``, Dipartimento di Fisica, P.le A.\\
Moro 2, I-00185 Roma, and INFM, Italy}
\author{M. Ferretti}
\address{Universit\`{a} di Genova, Dipartimento di Chimica e Chimica Fisica,\\
Via Dodecanneso 31, I-16146 Genova, and INFM, Italy}
\maketitle

\begin{abstract}
An increase of the acoustic absorption is found in La$_{2-x}$Sr$_{x}$CuO$%
_{4} $ ($x=0.019$, 0.03 and 0.06) close to the temperatures at which
freezing of the spin fluctuations in antiferromagnetic-correlated clusters
is expected to occur. The acoustic absorption is attributed to changes of
the sizes of the quasi-frozen clusters induced by the vibration stress
through magnetoelastic coupling.
\end{abstract}

%\draft
\twocolumn

\section{INTRODUCTION}

The low doping region of the phase diagram of La$_{2-x}$Sr$_{x}$CuO$_{4}$ is
attracting considerable interest, due to the appearance of unconventional
correlated spin dynamics and ordering processes (for a review see Ref.~%
\onlinecite{RBC98}). In undoped La$_{2}$CuO$_{4}$ the Cu$^{2+}$ spins order
into a 3D antiferromagnetic (AF) state with the staggered magnetization in
the $ab$ plane.\cite{VSM87} Doping by Sr rapidly destroys the long range AF
order, with $T_{N}$ passing from 315$~$K to practically 0$~$K around $%
x_{c}\simeq 0.02$. Above this critical value of the Sr content no long range
AF order is expected at finite temperature. There are also indications that
the holes are segregated into domain walls, sometimes identified as charge
stripes, which separate hole-poor regions where the AF correlations build up.%
\cite{CBJ92,BCC95} The holes should be mobile along these ''charge rivers'',
but at low $x$ they localize near the Sr atoms below $\sim 30~$K, causing a
distortion of the spin texture of the surrounding Cu$^{2+}$ atoms.\cite
{BCC95} For $x<x_{c}$ the spin distortions around the localized holes are
decoupled from the AF background, and freeze into a spin-glass (SG) state
below $T_{f}\left( x\right) \simeq \left( 815~\text{K}\right) x$. For $%
x>x_{c}$ a cluster spin-glass (CSG) state is argued to freeze below $%
T_{g}\left( x\right) \propto 1/x$ and AF correlations develop within the
domains defined by the charge walls, with the easy axes of the staggered
magnetization uncorrelated between different clusters. The formation of the
SG and CSG states are inferred from sharp maxima in the $^{139}$La NQR\cite
{CBJ92,RBC98,JBC99} and $\mu $SR\cite{NBB98} relaxation rates, which
indicate the slowing of the AF fluctuations below the measuring frequency ($%
\sim 10^{7}-10^{8}$~Hz in those experiments) on cooling, and from the
observation of irreversibility, remnant magnetization, and scaling behavior
in magnetic susceptibility experiments.\cite{CBK95,WUE99}

Here we report the observation of a step-like increase of the low-frequency
acoustic absorption close to the temperature at which the spin freezing
process is detected in the NQR measurements. The absorption is ascribed to
changes of the sizes of the frozen clusters induced by the vibration stress
through magnetoelastic coupling,\cite{NB} or equivalently to the motion of
the walls between them.

\section{EXPERIMENTAL AND RESULTS}

The samples where prepared by standard solid state reaction as described in
Ref.~\onlinecite{DFF94} and cut in bars approximately $40\times 4\times 0.6$%
\ mm$^{3}$. The final Sr contents and homogeneities where checked from the
temperature position and sharpness of the steps in the Young's modulus and
acoustic absorption due to the tetragonal (HTT) / orthorhombic (LTO)
transition, which occurs at a temperature $T_{t}$ linearly decreasing with
doping.\cite{Joh97} The transitions appear narrower in temperature than the
one of a Sr-free sample, indicating that the width was mostly intrinsic and
not due to Sr inhomogeneity, except for the sample at the lowest Sr content.%
\cite{76} The Sr concentrations estimated in this way turned out $%
x=0.0185\pm 0.0015$, $0.0315\pm 0.0015$ and $0.0645\pm 0.002$, in good
agreement with the nominal compositions. In the following the samples will
be referred as $x=0.019$, 0.03 and 0.06.

The complex Young's modulus $E$ was measured by electrostatically exciting
either of the lowest three flexural modes and detecting the vibration
amplitude by a frequency modulation technique. The elastic energy loss
coefficient (or reciprocal of mechanical $Q$) is related to the imaginary
part $E^{\prime \prime }$ of $E$ by $Q^{-1}\left( \omega ,T\right)
=E^{\prime \prime }\left( \omega ,T\right) /E^{\prime }\left( \omega
,T\right) $, and it was measured by the decay of the free oscillations or
the width of the resonance peak.

In Fig. 1 the anelastic spectra of three samples with $x=0.019$, 0.03, 0.06
below $16~$K measured exciting the first flexural mode are reported. A
step-like increase of the absorption is observed around or slightly below $%
T_{g}$ (Ref. \onlinecite{Joh97}) The gray arrows indicate the values of $%
T_{g}$ in the magnetic phase diagrams deduced from NQR\cite{CBJ92} (lower
values) and $\mu $SR\cite{NBB98} (higher values) experiments, which are in
agreement with the data in Ref. \onlinecite{Joh97} (for the sample with $%
x=0.019$ the $T_{g}\left( x=0.02\right) $ values are indicated). The black
arrows indicate the temperature of the maximum of the $^{139}$La NQR
relaxation rate measured on the same samples in a separate study,\cite{CCC00}
which indicate a freezing in the spin-glass phase, as discussed later. The
coincidence of the temperatures of the absorption steps with those of
freezing of the spin fluctuations suggest a correlation between the two
phenomena.

The sample with $x=0.03$ was outgassed from excess O by heating in vacuum up
to $790~$K, while the other two samples where in the as-prepared state,
therefore containing some interstitial O. The concentration $\delta $ of
excess O is a decreasing function of $x$ (Ref. \onlinecite{TR}) and should
be negligible for $x=0.06$ but not for $x=0.019$. This fact allowed us to
observe the absorption step singled out from the high-temperature tail of an
intense peak that occurs at lower temperature (see the sharp rise of
dissipation below 3$~$K for $x=0.019$ in Fig. 1). Such a peak has been
attributed to the tunneling-driven tilt motion of a fraction of the O
octahedra.\cite{76,61} The LTO phase is inhomogeneous on a local scale,\cite
{61,HSH96,BBK99,68} and a fraction of the octahedra would be unstable
between different tilt orientations, forming tunneling systems which cause
the anelastic relaxation process. The interstitial O atoms force the
surrounding octahedra into a fixed tilt orientation, resulting in a decrease
of the fraction of mobile octahedra and therefore in a depression of the
absorption peak. In addition, doping shifts the peak to lower temperature at
a very high rate, due to the coupling between the tilted octahedra and the
hole excitations.\cite{76} Therefore, it is possible to reduce the weight of
the low temperature peak by introducing concentrations of interstitial O\
atoms that are so small that do not change appreciably the doping level due
to the Sr substitutionals. Figure 2\ compares the absorption curves of the $%
x=0.019$ sample in the as-prepared state with a concentration $\delta \simeq
0.002$ of excess O and after removing it in vacuum at high temperature. The
initial concentration $\delta $ has been estimated from the intensity of the
anelastic relaxation process due to the hopping of interstitial O,\cite{63}
whose maximum occurs slightly below room temperature at our measuring
frequencies (not shown here). The presence of excess O indeed decreases and
shifts to lower temperature the tail of the peak in Fig.~2, while the effect
on the absorption step is negligible. This justifies the comparison of the
sample with $x=0.019$ and $\delta >0$ together with the other samples with $%
\delta \simeq 0$, and demonstrates that the nature of the low temperature
peak is different from that of the step-like absorption.

\section{DISCUSSION}

The present data show the presence of a step in the acoustic absorption at
the boundary of the spin-glass quasi-ordered state in the $T,x$ magnetic
phase diagram. The case of the $x=0.019$ sample is less clear-cut, since the
step is rather smooth. Furthermore, the Sr content is within the range $%
0.018<x<0.02$, at the boundary between the SG and the CSG phases, where the
phase diagram is largely uncertain.\cite{CCC00} The $T_{f}\left( x\right) $
line ends at $15~$K for $x\simeq 0.018$, and the line $T_{g}\left( x\right) $
starts from 10-12$~$K at $x\simeq 0.02$ (Refs. \onlinecite{WUE99,Joh97,CCC00}%
). A larger spread of experimental data\cite{Joh97} (from 7.8 to 12.5$~$K)
is actually observed just at $x=0.02$.

A\ mechanism which in principle produces acoustic absorption is the slowing
down of the magnetic fluctuations toward the spin-glass freezing. When
measuring the spectral density $J_{\text{spin}}\left( \omega ,T\right) $ of
the spin fluctuations (the Fourier transform of the spin-spin correlation
function), e.g. through the $^{139}$La NQR relaxation rate, a peak in $J_{%
\text{spin}}$ is found at the temperature at which the fluctuation rate $%
\tau ^{-1}\left( T\right) $ becomes equal to the measuring angular frequency 
$\omega $. Near the glass transition the magnetic fluctuation rate was found
to approximately follow the law\cite{CBJ92} $\tau ^{-1}\propto \left[ \left(
T-T_{g}\right) /T_{g}\right] ^{2}$, and the temperature at which the
condition $\omega \tau =1$ for the maximum of relaxation is satisfied for $%
\omega /2\pi =12-19$~MHz is close to $T_{g}$. A similar peak would be
observed in the spectral density of the lattice strain $J_{\text{latt}%
}\left( \omega ,T\right) $, if the spin fluctuations cause strain
fluctuations through magnetoelastic coupling. The acoustic absorption is
proportional to the spectral density of the strain and hence to $J_{\text{%
latt}}$, $Q^{-1}=\omega J_{\text{strain}}/T\propto \omega J_{\text{latt}}/T$%
, and therefore at our frequencies ($\omega \leq 50$~kHz) we should observe
a narrow peak at a temperature slightly lower than the ones detected by NQR
relaxation. The absorption steps in Fig.~1 can hardly be identified in a
strict way as due to the contribution from the freezing magnetic
fluctuations because they appear as steps instead of peaks. We propose that
the main contribution comes from the stress-induced movement of the domain
boundaries between the clusters of quasi-frozen antiferromagnetically
correlated spins. The mechanism is well known for ferromagnetic materials,%
\cite{NB} but is possible also for an ordered AF state, if an anisotropic
strain is coupled with the easy magnetization axis. In this case, the
elastic energy of domains with different orientations of the easy axis would
be differently affected by a shear stress, and the lower energy domains
would grow at the expenses of the higher energy ones. The dynamics of the
domain boundaries is different from that of the domain fluctuations and
generally produces broad peaks in the susceptibilities. An example is the
structural HTT/LTO transformation in the same samples, where the appearance
of the orthorhombic domains is accompanied by a step-like increase of the
acoustic absorption.\cite{LLN} We argue that the features in the anelastic
spectra just below $T_{g}$ are associated with the stress-induced motion of
the walls enclosing the clusters of AF correlated spins. More properly, the
anelastic relaxation is attributed to the stress-induced changes of the
sizes of the different domains.

The $x=0.019$ sample is at the border $x_{c}\simeq 0.02$ between SG and CSG
state. The NQR measurements on the same sample\cite{CCC00} indicate a
spin-freezing temperature $\sim 9~$K, closer to the CSG $T_{g}\left(
x_{c}\right) $ rather then to the SG $T_{f}\left( x_{c}\right) $, which is
consistent with the presence of moving walls, otherwise absent in the SG
state. Nonetheless, following the model proposed by Gooding {\it et al.}\cite
{GSB97} we do not expect a sharp transition between the SG and the CSG
states. According to that model, at low temperature the holes localize near
the Sr dopants, and in the ground state an isolated hole circulates
clockwise or anti-clockwise over the four Cu atoms neighbors to Sr. Such a
state induces a distortion of the surrounding Cu spins, otherwise aligned
according to the prevalent AF order parameter. The spin texture arising from
the frustrated combination of the spin distortions from the various
localized holes produces domains with differently oriented AF order
parameters, which can be identified with the frozen AF spin clusters. The
dissipative dynamics which we observe in the acoustic response should arise
from the fact that the energy surface of the possible spin textures has many
closely spaced minima\cite{GSB97} and the vibration stress, through
magnetoelastic coupling, can favor jumps to different minima. In this
picture, one could argue that the random distribution of Sr atoms may cause
the formation of spin clusters also for $x\lesssim x_{c}$ and it is possible
to justify the fact that for $x=0.019$ the absorption step does not start
below the maximum of the $^{139}$La NQR\ relaxation rate, which signals the
freezing of the spin clusters. Rather, the acoustic absorption slowly starts
increasing slightly before the $T_{g}$ determined by the NQR maximum is
reached. This may indicate that the spin dynamics is not only governed by
cooperative freezing, but is also determined by the local interaction with
the holes localized at the surrounding Sr atoms. Then, the regions in which
the Sr atoms induce a particularly strong spin-texture could freeze and
cause anelastic relaxation before the cooperative transition to the glass
state is completed. Systematic measurements around the $x=0.02$ doping range
are necessary to clarify this point.

The dependence of the intensity of the absorption step on $x$, which is
sharper and most intense at $x=0.03$, qualitatively supports the above
picture. In fact, at lower doping one has only few domains embedded in a
long range ordered AF background, while above $0.05$ the fraction of walls
of disordered spins connecting the Sr atoms increases at the expenses of the
ordered domains, with a cross-over to incommensurate spin correlations.\cite
{GSB97} The anelasticity due to the stress-induced change of the domain
sizes is expected to be strongest in correspondence to the greatest fraction
of ordered spins, namely between $0.03$ and $0.05$, in accordance with the
spectra in Fig.~1.

Finally we point out the insensitiveness of the absorption step to the
presence of interstitial O (Fig. 2 and Ref. \onlinecite{76}), in view of the
marked effects that even small quantities of excess O cause to the low
temperature peak (Fig. 2) and to the rest of the anelastic spectrum.\cite
{61,76} This is consistent with a dissipation mechanism of magnetic rather
than of structural origin.

\section{CONCLUSION}

The elastic energy loss coefficient of La$_{2-x}$Sr$_{x}$CuO$_{4}$
(proportional to the imaginary part of the elastic susceptibility) measured
around $10^{3}$~Hz in samples with $x=0.019$, 0.032 and 0.064 shows a
step-like rise below the temperature of the transition to a quasi-frozen
cluster spin-glass state. The origin of the acoustic absorption is thought
to be magnetoelastic coupling, namely anisotropic in-plane strain associated
with the direction of the local staggered magnetization. The absorption is
not peaked at $T_{g}$ and therefore does not directly correspond to the peak
in the dynamic spin susceptibility due to the spin freezing. Rather, it has
been ascribed to the stress-induced changes of the sizes of the spin
clusters, or equivalently to the motion of the walls. The phenomenology is
qualitatively accounted for in the light of the model of Gooding {\it et al.}%
\cite{GSB97} of magnetic correlations of the Cu$^{2+}$ spins induced by the
holes localized near the Sr dopants.

\section*{Acknowledgments}

The authors thank Prof. A. Rigamonti for useful discussions and for a
critical review of the manuscript. This work has been done in the framework
of the Advanced Research Project SPIS of INFM.

%\bibliographystyle{unsrt}
%\bibliography{altro,ccc,htcsc}

\begin{references}
\bibitem{RBC98}  A. Rigamonti, F. Borsa and P. Carretta, Rep. Prog. Phys. 
{\bf 61}, 1367 (1998).

\bibitem{VSM87}  D. Vaknin, S.K. Sinha, D.E. Moncton, D.C. Johnston, J.M.
Newsam, C.R. Safinya and H.E. King Jr., Phys. Rev. Lett. {\bf 58}, 2802
(1987).

\bibitem{CBJ92}  J.H. Cho, F. Borsa, D.C. Johnston and D.R. Torgeson, Phys.
Rev. B {\bf 46}, 3179 (1992).

\bibitem{BCC95}  F. Borsa, P. Carretta, J.H. Cho, F.C. Chou, Q. Hu, D.C.
Johnston, A. Lascialfari, D.R. Torgeson, R.J. Gooding, N.M. Salem and K.J.E.
Vos, Phys. Rev. B {\bf 52}, 7334 (1995).

\bibitem{JBC99}  M.-H. Julien, F. Borsa, P. Carretta, M. Horvati, C.
Berthier and C.T. Lin, Phys. Rev. Lett. {\bf 83}, 604 (1999).

\bibitem{NBB98}  Ch. Niedermayer, C. Bernhard, T. Blasius, A. Golnik, A.
Moodenbaugh and J.I. Budnick, Phys. Rev. Lett. {\bf 80}, 3843 (1998).

\bibitem{CBK95}  F.C. Chou, N.R. Belk, M.A. Kastner, R.J. Birgeneau and A.
Aharony, Phys. Rev. Lett. {\bf 75}, 2204 (1995).

\bibitem{WUE99}  S. Wakimoto, S. Ueki, Y. Endoh and K. Yamada,
cond-mat/9910400.

\bibitem{NB}  A.S. Nowick and B.S. Berry, {\it Anelastic Relaxation in
Crystalline Solids}. (Academic Press, New York, 1972).

\bibitem{DFF94}  M. Daturi, M. Ferretti and E.A. Franceschi, Physica C {\bf %
235-240}, 347 (1994).

\bibitem{Joh97}  D.C. Johnston, {\it Handbook of Magnetic Materials}. ed. by
K.H.J. Buschow, p. 1 (North Holland, 1997).

\bibitem{76}  F. Cordero, R. Cantelli and M. Ferretti, cond-mat/9910402, to
be published in Phys. Rev. B.

\bibitem{CCC00}  A. Campana, R. Cantelli, M. Corti, F. Cordero and A.
Rigamonti, unpublished.

\bibitem{TR}  E. Takayama-Muromachi and D.E. Rice, Physica C {\bf 177}, 195
(1991).

\bibitem{61}  F. Cordero, C.R. Grandini, G. Cannelli, R. Cantelli, F.
Trequattrini and M. Ferretti, Phys. Rev. B {\bf 57}, 8580 (1998).

\bibitem{HSH96}  D. Haskel, E.A. Stern, D.G. Hinks, A.W. Mitchell, J.D.
Jorgensen and J.I. Budnick, Phys. Rev. Lett. {\bf 76}, 439 (1996).

\bibitem{BBK99}  E.S. Bozin, S.J.L. Billinge, G.H. Kwei and H. Takagi, Phys.
Rev. B {\bf 59}, 4445 (1999).

\bibitem{68}  F. Cordero, R. Cantelli, M. Corti, M. Campana and A.
Rigamonti, Phys. Rev. B {\bf 59}, 12078 (1999).

\bibitem{63}  F. Cordero, C.R. Grandini and R. Cantelli, Physica C {\bf 305}%
, 251 (1998).

\bibitem{LLN}  W.-K. Lee, M. Lew and A.S. Nowick, Phys. Rev. B {\bf 41}, 149
(1990).

\bibitem{GSB97}  R.J. Gooding, N.M. Salem, R.J. Birgeneau and F.C. Chou,
Phys. Rev. B {\bf 55}, 6360 (1997); K.S.D. Beach and R.J. Gooding,
cond-mat/0001095.
\end{references}

\section{Figures captions}

Fig. 1 Elastic energy loss coefficient of La$_{2-x}$Sr$_{x}$CuO$_{4}$ with $%
x=0.019$ ($1.3$~kHz), $x=0.03$ ($1.7$~kHz), $x=0.06$ ($0.43$~kHz). The gray
arrows indicate the temperature $T_{g}$ of freezing into the cluster
spin-glass state deduced from NQR\cite{CBJ92} (lower values) and $\mu $SR%
\cite{NBB98} (higher values) experiments The black arrows indicate the
temperature of the maximum of the $^{139}$La NQR relaxation rate measured on
the same samples.\cite{CCC00}

Fig. 2 Elastic energy loss coefficient of the sample with $x=0.019$ in the
as prepared state (with interstitial O) and after outgassing the excess O
measured at 1.3~kHz.

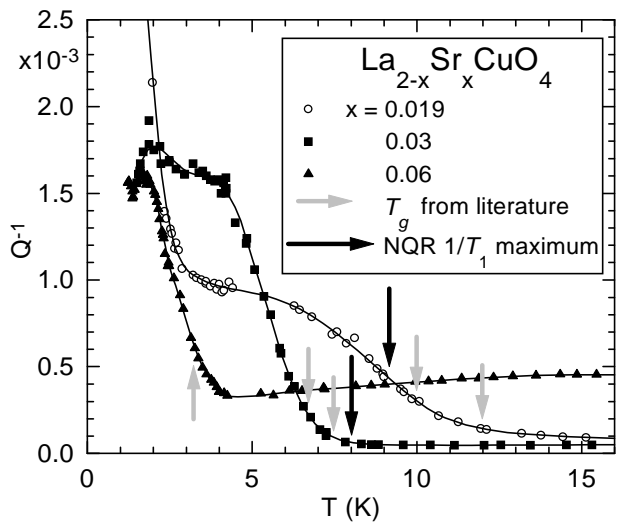
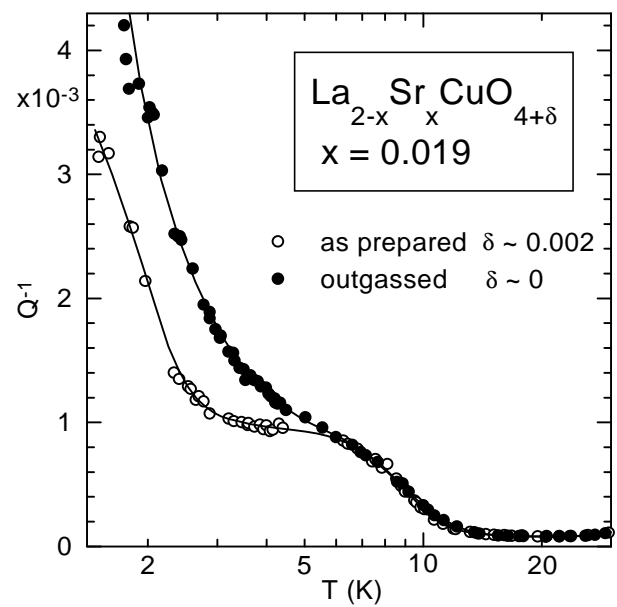
\end{document}